\pdfoutput=1
\documentclass[sigconf]{acmart}

\acmConference[ICSE 2024]{46th International Conference on Software Engineering}{April 2024}{Lisbon, Portugal}

\AtBeginDocument{%
  }

\copyrightyear{2024}
\acmYear{2024}
\setcopyright{acmlicensed}\acmConference[ICSE '24]{2024 IEEE/ACM 46th
International Conference on Software Engineering}{April 14--20, 2024}{Lisbon,
Portugal}
\acmBooktitle{2024 IEEE/ACM 46th International Conference on Software
Engineering (ICSE '24), April 14--20, 2024, Lisbon, Portugal}
\acmDOI{10.1145/3597503.3639077}
\acmISBN{979-8-4007-0217-4/24/04}

\usepackage{framed}
\usepackage{listings}
\usepackage{xcolor}
\usepackage{color}
\usepackage{multirow}
\usepackage{algorithm}

\usepackage{amsmath}
\usepackage{amssymb}
\usepackage{threeparttable}
\usepackage{url}
\usepackage[most]{tcolorbox}
\usepackage{hyperref}
\usepackage[hyphenbreaks]{breakurl}
\usepackage{algpseudocode}
\usepackage{colortbl}
\hypersetup{
    colorlinks = true,
    linkcolor=blue,
    filecolor=blue,      
    urlcolor=blue,
    citecolor=cyan,
}
\usepackage{xspace}
\usepackage{subfigure}
\usepackage{graphicx}
\usepackage{fvextra}
\usepackage{pifont}

\usepackage[shortlabels]{enumitem}
\setlist[enumerate]{nosep}

\definecolor{codegreen}{rgb}{0,0.6,0}
\definecolor{codegray}{rgb}{0.5,0.5,0.5}
\definecolor{codepurple}{rgb}{0.58,0,0.82}
\definecolor{backcolour}{rgb}{0.95,0.95,0.92}
\definecolor{lightgreen}{rgb}{0,0.4,0}
\definecolor{lightred}{rgb}{0.4,0,0}

\definecolor{mygray}{gray}{.9}

\lstdefinestyle{mystyle}{
    commentstyle=\color{brown},
    keywordstyle=\color{magenta},
    numberstyle=\tiny\color{codegray},
    stringstyle=\color{codepurple},
    basicstyle=\ttfamily\footnotesize,
    breakatwhitespace=false,         
    breaklines=true,                 
    captionpos=b,                    
    keepspaces=true,                 
    numbers=left,                    
    numbersep=5pt,                  
    showspaces=false,                
    showstringspaces=false,
    showtabs=false,                  
    tabsize=2,
    escapeinside={<@}{@>},
    language=python
}

\newcommand{\tool}{\textsc{PyConf}\xspace}
\newcommand{\bench}{\textsc{VLibs}\xspace}

\newcommand\etal{{\it{et al.\ }}}

\newcommand{\tabincell}[2]{\begin{tabular}{@{}#1@{}}#2\end{tabular}}
\newcommand{\eat}[1]{\if 0 #1 \fi}

\newfont{\mycrnotice}{ptmr8t at 7pt}
\newfont{\myconfname}{ptmri8t at 7pt}

\newcommand{\finding}[2]{
\begin{tcolorbox}[breakable,width=\linewidth,boxrule=0pt,top=1pt, bottom=1pt, left=1pt,right=1pt, colback=gray!20,colframe=gray!20]
\textbf{Finding #1:} #2
\end{tcolorbox}
}

\begin{document}

\title{Less is More? An Empirical Study on Configuration Issues in Python PyPI Ecosystem}

\author{Yun Peng}
 \affiliation{\institution{The Chinese University of Hong Kong}\country{Hong Kong, China}}
 \email{ypeng@cse.cuhk.edu.hk}

\author{Ruida Hu}
 \affiliation{\institution{Harbin Institute of Technology} \country{Shenzhen, China}}
 \email{200111107@stu.hit.edu.cn}

 \author{Ruoke Wang}
 \affiliation{\institution{Harbin Institute of Technology} \country{Shenzhen, China}}
 \email{200110930@stu.hit.edu.cn}

\author{Cuiyun Gao}
 \affiliation{\institution{Harbin Institute of Technology} \country{Shenzhen, China}}
 \email{gaocuiyun@hit.edu.cn}
 \authornote{Corresponding author}

\author{Shuqing Li}
 \affiliation{\institution{The Chinese University of Hong Kong}\country{Hong Kong, China}}
 \email{sqli21@cse.cuhk.edu.hk}

\author{Michael R. Lyu}
 \affiliation{\institution{The Chinese University of Hong Kong} \country{Hong Kong, China}}
 \email{lyu@cse.cuhk.edu.hk}


\begin{abstract}
Python is the top popular programming language used in the open-source community, largely owing to the extensive support from diverse third-party libraries within the PyPI ecosystem. Nevertheless, the utilization of third-party libraries can potentially lead to conflicts in dependencies, prompting researchers to develop dependency conflict detectors. Moreover, endeavors have been made to automatically infer dependencies. These approaches focus on version-level checks and inference, based on the assumption that configurations of libraries in the PyPI ecosystem are correct. However, our study reveals that this assumption is not universally valid, and relying solely on version-level checks proves inadequate in ensuring compatible run-time environments.

In this paper, we conduct an empirical study to comprehensively study the configuration issues in the PyPI ecosystem. Specifically, we propose \tool, a source-level detector, for detecting potential configuration issues. \tool employs three distinct checks, targeting the setup, packing, and usage stages of libraries, respectively. To evaluate the effectiveness of the current automatic dependency inference approaches, we build a benchmark called \bench, comprising library releases that pass all three checks of \tool. We identify 15 kinds of configuration issues and find that 183,864 library releases suffer from potential configuration issues. Remarkably, 68\% of these issues can only be detected via the source-level check. Our experiment results show that the most advanced automatic dependency inference approach, PyEGo, can successfully infer dependencies for only 65\% of library releases. The primary failures stem from dependency conflicts and the absence of required libraries in the generated configurations. Based on the empirical results, we derive six findings and draw two implications for open-source developers and future research in automatic dependency inference.
\end{abstract}



\maketitle

\section{Introduction}\label{sec:intro}

\begin{figure}[t]
    \centering
    \includegraphics[width = 0.47\textwidth]{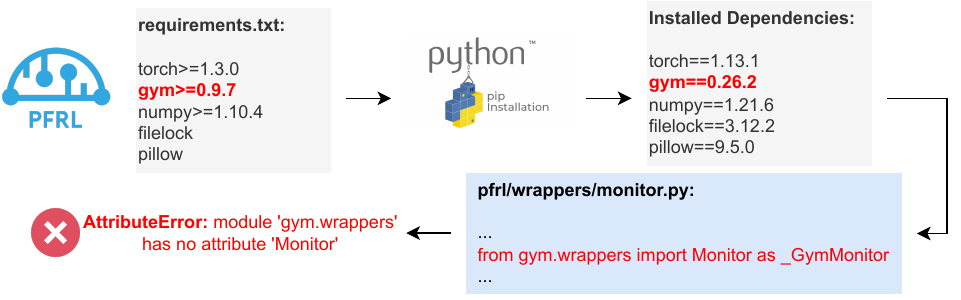}
    \caption{A configuration issue of the third-party library PFRL.}
    \label{fig:mot}
\end{figure}

Python has experienced a remarkable 22.5\% year-over-year surge in usage, positioning it as the second most favored programming language within the GitHub open-source community~\cite{octoverse}. The popularity of Python is primarily established by its flexible and readable syntax, making it easier for developers to maintain complicated software. Nowadays, the success of Python owes much to its thriving and supportive community, which plays a pivotal role in fostering its prosperity. The accessibility and utility of Python are further amplified by the public libraries available on the Python Package Index (PyPI) platform. With over 470 thousand Python projects and more than 4.7 million releases~\cite{pypi}, PyPI serves as the primary repository for numerous third-party libraries. By encapsulating reusable functionalities with APIs in third-party libraries, developers can easily build complicated applications.

In the dynamic ecosystem of third-party libraries hosted on PyPI, multiple releases of the same library are often available, distinguished by version numbers. To use a specific library release, developers must specify both the library's name and the desired version. Utilizing the official library management tool, \texttt{pip}~\cite{pip}, for PyPI, developers can effortlessly retrieve and install the intended release based on the associated configurations. Once installed in the current run-time environment, the library release can be accessed through import statements within the source code. Compared with static programming languages such as Java and C/C++, third-party library usage in Python is much simpler and requires no compilation. However, even with this streamlined approach, the presence of any configuration issues in the third-party libraries can lead to potential run-time failures.

Numerous research efforts~\cite{artho12why,jibesh18conflictjs,sotovalero19the,wang18do,wang19could,wang20watchman} are dedicated to detecting potential dependency conflicts among diverse third-party libraries during the constraint-solving process. As a library may rely on others, a dependency graph can be established to represent the interconnected libraries with nodes and version constraints with edges. Based on the dependency graph, these approaches use SMT solvers to determine an available version assignment for each library. In addition, some other work~\cite{horton19dockerizeme,cheng22conflict,ye22knowledge} develops knowledge graphs for third-party libraries on PyPI and then builds run-time environments for new Python projects based on the knowledge graphs.

The aforementioned version-level approaches have been established and evaluated under the assumption that the configurations of existing Python projects are accurate, as they solely examine version constraints in configurations without inspecting the source code. However, we have discovered instances where this assumption does not hold, and we present an illustrative example in Fig.~\ref{fig:mot}. In this example, the third-party library \texttt{PFRL}~\cite{pfrl} implements several well-known reinforcement learning algorithms. It records all required third-party libraries in the \texttt{requirements.txt} file. During installation, the library manager \texttt{pip} resolves the constraints in \texttt{requirements.txt} and installs the latest available version for each library. A widely-used library, \texttt{gym}, is among the dependencies specified in \texttt{requirements.txt}. The configuration for \texttt{gym} merely requires a version newer than 0.9.7. As a result, \texttt{pip} installs the latest version, 0.26.2, into the project as it satisfies the constraint\footnote{The installation was performed in July 2023.}. Since version 0.26.2 of \texttt{gym} does not conflict with other libraries in \texttt{requirements.txt}, it passes the regular conflict check and becomes part of the run-time environment. However, when running the code in \texttt{PFRL}, an \texttt{AttributeError} is raised as the \texttt{Monitor} class from \texttt{gym.wrappers} in the file \texttt{pfrl/wrappers/monitor.py} cannot be found. This issue is widely discussed on \texttt{PFRL}'s GitHub issues~\cite{issue} and Stack Overflow~\cite{stackoverflow}. The root cause is that \texttt{gym} removed the \texttt{Monitor} class starting from version 0.23.0. Since this change only affects the source code and is not detected by version-level checks, the problem remains unnoticed. This scenario highlights the inadequacy of version-level checks in ensuring the compatibility of source code and run-time environments. To address this problem, the library \texttt{gym} should be constrained to versions \texttt{gym>=0.9.7, gym<0.23.0}. However, predicting such changes in \texttt{gym} during the development of \texttt{PFRL} is not feasible since version 0.23.0 of \texttt{gym} had not been released at that time. Therefore, the configurations in Python projects can be outdated despite being correct at the release time.

The above-mentioned challenge of version-level dependency checks may pose big threats to the development and evaluation of automatic dependency inference approaches that heavily depend on PyPI library configurations. To address this challenge, we first comprehensively study the potential configuration issues in the PyPI ecosystem (RQ1) and then construct a source-level compatible dataset to facilitate the evaluation of existing automatic dependency inference approaches (RQ2).

To answer RQ1, we introduce \tool, an automatic approach designed to identify both version-level and source-level configuration issues in third-party libraries on the PyPI platform. \tool incorporates three distinct checks, namely \textit{Installation Check}, \textit{Dependency Check} and \textit{Import Validation}, to detect configuration issues during the setup stage, packing stage and usage stage of third-party libraries, respectively. Through an analysis of \tool's results, we identify 183,864 (54\%) library releases among the 338,069 checked releases that exhibit potential configuration issues. Notably, 68\% of these issues are newly detected by the source-level check, i.e., the \textit{Import Validation}. 
We identify 15 kinds of configuration issues based on the run-time error types and classify them into three major categories: \textit{Incomplete Configuration}, \textit{Incorrect Configuration} and \textit{Incorrect Code}. For RQ2, we construct a benchmark, \bench, consisting of 131,720 library releases that successfully pass all three checks implemented by \tool. We then evaluate the correctness of the inferred run-time environments by the three state-of-the-art automatic dependency inference approaches Pipreqs~\cite{pipreqs}, Dockerizeme~\cite{horton19dockerizeme} and PyEGo~\cite{ye22knowledge}, respectively.

\textbf{Key Findings.} 
Based on a thorough analysis of the experiment results pertaining to RQ1 and RQ2, we have summarized the following key findings:

    1) Developers tend to provide inadequate configurations for the usage of libraries, especially for Python versions and direct imports in source code.
    
    2) Developers make mistakes in writing configurations since 19\% of configuration issues are incorrect configurations. What's more, about 50\% incorrect configuration issues can only be detected by \textit{Import Validation}, indicating the importance of source-level validation.
    
    3) Current automatic dependency inference approaches fail to infer about 35\% of Python projects. Among the failures, the majority are attributed to dependency conflicts and the absence of required libraries in the generated configurations.

\begin{figure*}[t]
    \centering
    \includegraphics[width = 0.9\textwidth]{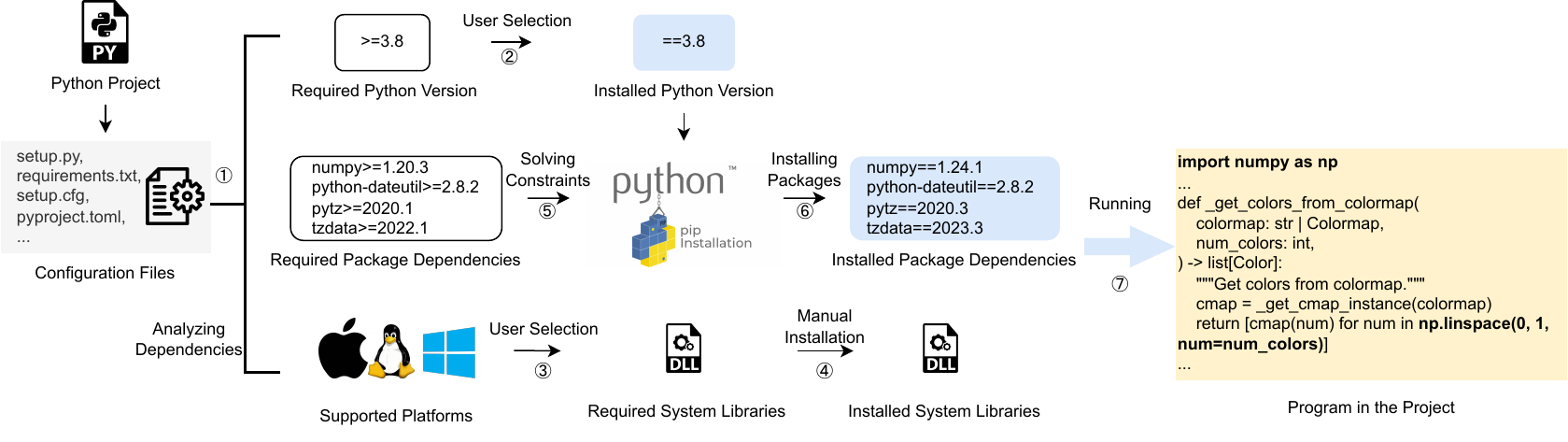}
    \caption{The typical process of run-time environment installation for Python projects.}
    \label{fig:background}
\end{figure*}

Based on the findings, we conclude two implications for the developers of third-party libraries on the PyPI platform and the future research of automatic dependency inference. Specifically, we find that ``\textit{less is more}'', i.e., fewer dependency constraints can lead to more configuration errors, so we
suggest developers avoid employing open constraints such as \texttt{version>1.0}, but set complete and strict dependency constraints limiting the versions of dependencies to the verified ones before the release dates. For future research on automatic dependency inference, we suggest researchers add more conflict checks to avoid generating incorrect configurations.

\textbf{Contributions.} To sum up, we list our contributions as follows.
\begin{itemize}
    \item To the best of our knowledge, we are the first to study the source-level configuration issues in the PyPI ecosystem systematically. 
    \item We propose an automatic approach \tool that incorporates \textit{Installation Check}, \textit{Dependency Check} and \textit{Import Validation} to detect configuration issues for Python projects.
    \item We build a benchmark \bench that includes
    131,720 library releases to facilitate the evaluation of automatic dependency inference approaches.
\end{itemize}
\section{Python Run-time Environment}\label{sec:background}

Python, as an interpreted programming language, offers the advantage of not requiring compilation prior to execution.  This attribute facilitates fast prototyping and enables Python programs to be executed on different platforms. However, benefiting from rich support from external libraries, nearly all Python projects depend on multiple third-party or system libraries to avoid redundant implementations of common functionalities.  Therefore, it becomes imperative to establish the appropriate run-time environment, comprising all necessary libraries, before running a Python project effectively.

We illustrate the process of installing the run-time environment for a Python project based on its source code in Fig.\ref{fig:background}. Typically, developers document all project dependencies in configuration files. As the Python community evolves, various configuration formats like \texttt{requirements.txt} and \texttt{setup.py} have emerged. Additionally, there are diverse developer tools, such as \texttt{setuptools}, available to analyze these configuration files (\textcircled{1} in Fig.\ref{fig:background}). The configuration files contain three types of dependencies: 1) the required Python version, 2) the necessary third-party libraries, and 3) the required platform and corresponding system libraries.

In most cases, users must first select the appropriate Python version and platform (\textcircled{2} and \textcircled{3} in Fig.\ref{fig:background}) before proceeding with the installation of other dependencies. If the Python project relies on some system libraries of the selected platform, users may also need to install them manually (\textcircled{4} in Fig.~\ref{fig:background}). Since Python third-party libraries are hosted on the PyPI platform~\cite{pypi}, Python Software Foundation also provides a dedicated tool named \texttt{pip}~\cite{pip} to facilitate automated installation. \texttt{Pip} first resolves the constraints of third-party libraries provided in the configuration file (\textcircled{5} in Fig.~\ref{fig:background}), and then selects the latest valid version for each library (\textcircled{6} in Fig.~\ref{fig:background}). By ensuring the presence of the appropriate Python version, third-party libraries, and system libraries, users can successfully execute certain Python projects (\textcircled{7} in Fig.~\ref{fig:background}).

\section{Methodology}\label{sec:approach}

In this section, we introduce how we collect the metadata of PyPI libraries and how \tool works to detect potential configuration issues.

\begin{figure*}[t]
    \centering
    \includegraphics[width = 0.8\textwidth]{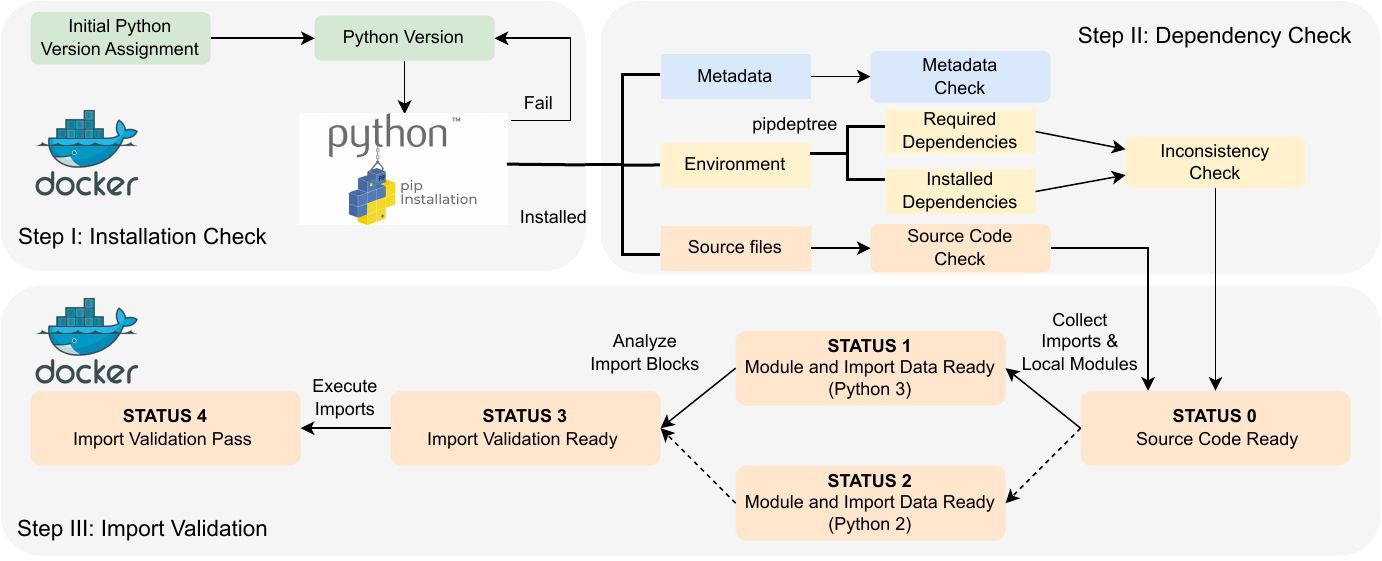}
    \caption{The overview of \tool.}
    \label{fig:overview}
\end{figure*}

\subsection{Data Preparation}

As of July 2023, the PyPI ecosystem boasts a substantial collection of approximately 471,000 libraries, encompassing over 4,712,000 releases~\cite{pypi}. It is quite difficult to perform a comprehensive analysis of all the libraries and their releases on a single machine. To address this challenge, we employ a well-established strategy used in prior studies~\cite{horton19dockerizeme,ye22knowledge,cheng22conflict} and collect data from the top 10,000 most popular libraries, as reported by libraries.io~\cite{librariesio}. Libraries with only one or two releases, which generally do not necessitate automatic version determination, are excluded from our analysis, resulting in a dataset comprising 8,282 libraries and 338,069 releases\footnote{Data was collected in November 2022.}. The first column of Table~\ref{tab:statistics} provides statistics on these libraries. In accordance with Python Enhancement Proposal (PEP) 508~\cite{pep508}, names of PyPI libraries are case-insensitive, and distinctions between dash, dot, and underscore are disregarded. To ensure consistency and avoid multiple names for the same library, we normalize all library names to lowercase and replace all dots and underscores with dashes.

\textbf{Initial Python Version Assignment.} As mentioned in Sec.~\ref{sec:background}, a smooth \texttt{pip} installation requires the correct Python version. Hence, we begin by assigning an initial Python version to each library release in the dataset. To acquire the Python version constraints for each library release, we examine the classifiers set by developers on the project web page of the PyPI platform. PyPI offers a set of classifiers for developers to denote the compatibility status of library releases. Among these classifiers, those categorized under the programming language category specify the Python versions with which a library release is compatible.
For instance, the developers of library release \texttt{pipreqs-0.4.13} add classifier \texttt{Python::3.7} in the web page~\cite{pipreqsweb}, indicating that \texttt{pipreqs-0.4.13} can be used in Python version 3.7. By collecting such classifiers from the web pages, we determine the latest Python version applicable to each library release as the initial Python version.

The initial Python versions inferred from classifiers provide relatively reliable insights into developers' intentions regarding library usage. However, setting classifiers is not mandatory when developers publish a new release on PyPI, so we cannot assign initial Python versions for certain library releases lacking appropriate classifiers. To tackle this problem, we collect the release dates of such library releases and select the latest Python version released 180 days before the release dates of the library releases. This assignment may not be accurate but can be fixed by \tool when the installation fails.

\subsection{\tool: Detecting Configuration Issues}

\begin{table}[t]
    \centering
    \caption{The statistics of the PyPI libraries in our study. ``Installed'' and ``Validated'' indicates the libraries passing the Dependency Check and all checks of \tool, respectively. \#Stars indicate the number of GitHub Stars of libraries. \#Stars, \#Classes, \#Functions and \#Imports are shown in the format of Avg/Max/Min. The data of \#Stars is calculated per library and others are calculated per release. Note that the source code data in the first column is not available as the libraries are not installed.}
    \scalebox{0.9}{
    \begin{tabular}{lccc}
    \toprule
    & \textbf{All} & \textbf{Installed} & \textbf{Validated} (\textsc{VLibs}) \\
    \midrule
    \#Libraries & 8,282 & 7,830 & 5,371 \\
    \#Releases & 338,069 & 303,377& 131,720 \\
    \#Modules & - & 368,304 & 144,250 \\
    \#Stars (k) & 2.3/159.0/0.0 & - & -  \\
    \#Classes (k)  & - & 0.3/88.1/0.0  & 0.2/21.9/0.0 \\
    \#Functions (k)  & - & 1.5/261.4/0.0 & 0.6/50.3/0.0\\
    \#External Imports  & - & 49/2207/0 & 23/386/0\\
    \#Lines of Code (k) &- & 18.2/7455.3/0.0 & 6.5/551.3/0.0 \\
    \bottomrule
    \end{tabular}}
    \label{tab:statistics}
\end{table}

\tool checks both version-level and source-level configuration issues for libraries in the PyPI ecosystem. We present the overview of \tool in Fig.~\ref{fig:overview}. \tool conducts three checks, namely \textit{Installation Check}, \textit{Dependency Check} and \textit{Import Validation}, to discover potential configuration issues in the setup stage, the packing stage and the usage stage of libraries, respectively. The \textit{Installation Check} verifies the availability of the library releases and detects fatal configuration errors, such as dependency conflicts, that even prevent successful library installation. The \textit{Dependency Check} verifies the consistency of the installed environment with the specified configuration, correctness of the library metadata and syntactic correctness of the source code, which are threatened by mistakes made during the packing stage before a library is published. The \textit{Import Validation} verifies the compatibility of the source code with the installed run-time environment to discover run-time errors during the usage of libraries.

\textbf{Installation Check.} Upon receiving the name and version of a library, \tool initiates an empty run-time environment within a docker container using the initial Python version. The library release is then installed using the command \texttt{pip install <library> == <version>}. However, due to certain initial Python version assignments being estimated based on release dates, and the existence of erroneous configurations authored by developers, some library releases may fail to be installed under the initial Python version. For libraries with Python version constraints, \tool retries the installation using another valid Python version. For libraries lacking such constraints or failing on all versions indicated by the constraints, \tool adopts a heuristic searching approach to minimize overhead. Specifically, \tool first copies the Python version of other successfully installed releases of the same library as different releases of the same library require similar run-time environments. In cases where the installation still fails, \tool attempts commonly used versions such as 2.7, 3.6, and 3.10. The heuristic searching strategy can handle most installation failures and \tool resorts to trying all possible Python versions only when the heuristic search proves unsuccessful. Therefore, the installation check fails only when there is no compatible Python version for the given library release or when there are critical errors in applying the configurations provided by developers. This indicates that the library release is not available for use under any Python version.

\textbf{Dependency Check.} 
During the installation process of a library release via \texttt{pip}, three types of data are downloaded into the system:

1) \textit{Metadata}. The metadata is stored in the format of a folder named \texttt{<package>-<version>.dist-info}. To analyze this metadata, \tool focuses on the \texttt{top\_level.txt} file, which enumerates all modules that can be imported from the library release.

2) \textit{Run-time environment.} \tool captures information regarding the installed run-time environment, including the versions of installed third-party libraries and the version constraints of the required third-party libraries, via the \texttt{pipdeptree}~\cite{pipdeptree} tool. \tool then proceeds to resolve the version constraints of the required third-party libraries and cross-checks them against the installed versions to detect potential inconsistencies.

3) \textit{Source files.} To validate the syntactic correctness of the source code, \tool locates source folders or files based on the modules collected from the metadata. It employs the \texttt{ast} module~\cite{ast} to parse all source files and identifies the presence of any syntax errors.

\begin{algorithm}[t]
\caption{Import Block Analysis}
\begin{algorithmic}[1]
\Require
Abstract Syntax Tree (AST) of the current source file, $ast$;
\Ensure
Import blocks, $B$; Block-free Imports, $D$;
\Function{getImportBlocks}{$block$}\Comment{The main function}
\State $importBlocks \leftarrow$ \{\}; $subBlocks \leftarrow$ divideBlock($block$)
\For{$sb \in subBlocks$}
\State $curIB \leftarrow$ \{\}; $curBFI \leftarrow$ \{\}
\For{$node \in sb$.importnodes}
\If{isIforTryOutside($node$)}
\State $bNode \leftarrow$ getOutmostIforTryNode($node$)
\State $curIB \leftarrow curIB$ + \{\Call{getImportBlocks}{$bNode$}\}
\Else
\State $curBFI\leftarrow$ $curBFI$ + \{$node$\}
\EndIf
\EndFor
\State $curB\leftarrow curIB$ + \{$curBFI$\}
\State $importBlocks \leftarrow importBlocks$ + \{$curB$\}
\EndFor
\State \Return $importBlocks$
\EndFunction

\State $blocks$ $\leftarrow$ \{ \}; $D$ $\leftarrow$ \{ \}; $B$ $\leftarrow$ \{ \}\Comment{The overall algorithm}
\For{$node \in ast$.importnodes}
\If{isIforTryOutside($node$)}
\State $blocks \leftarrow blocks$ + \{getOutmostIforTryNode($node$)\}
\Else
\State $D \leftarrow D$ + \{$node$\}  
\EndIf
\EndFor
\For{$b \in blocks$}
\State $B \leftarrow B$ + \{\Call{getImportBlocks}{$b$}\}
\EndFor

\end{algorithmic}
\label{alg:blockanalysis}
\end{algorithm}

\textbf{Import Validation.} Successful installation and consistent run-time environment do not necessarily guarantee the smooth usage of the library, since the execution still fails if some external import requirements in the source code cannot be fulfilled. \tool conducts \textit{Import Validation} to detect these issues. \tool leverages a finite state machine (FSM) with four states to guide the process of \textit{Import Validation}, as shown in step III of Fig.~\ref{fig:overview}. 

\textit{1) Collect Imports and Local Modules (STATUS 0 $\rightarrow$ STATUS 1/2).} Initially, all library releases enter STATUS 0 if \tool can successfully locate their source code in \textit{Dependency Check}. For library releases with STATUS 0, \tool collects import statements in the source code.
Import statements in the source code can be of two types: \textit{internal imports}, which introduce local modules within the project, and \textit{external imports}, which require third-party libraries from the run-time environment.
\tool employs different approaches to handle the two kinds of import statements. 

Local modules are required to distinguish internal imports and external imports. Different source files also have different available local modules. For each source file in the library release, \tool collects the names of all Python source files and the sub-directories with \texttt{\_\_init\_\_.py} file in the same directory, as well as image files such as \texttt{.so} and \texttt{.pyd}, as local modules. Next, \tool checks all import statements in the source file and compares the imported module with the local modules to identify internal imports. Since internal imports are not pertinent to the run-time environment, they are excluded from the \textit{Import Validation} process. The remaining import statements are regarded as external imports. \tool executes external imports in the installed run-time environment to detect potential compatibility issues. 

If the above process succeeds under Python 3, the library release enters STATUS 1. Otherwise, \tool retries the similar process under Python 2 with some small adaptations to Python 2 syntax. The library release analyzed under Python 2 enters STATUS 2.

\begin{figure}[t]
    \centering
    \includegraphics[width = 0.46\textwidth]{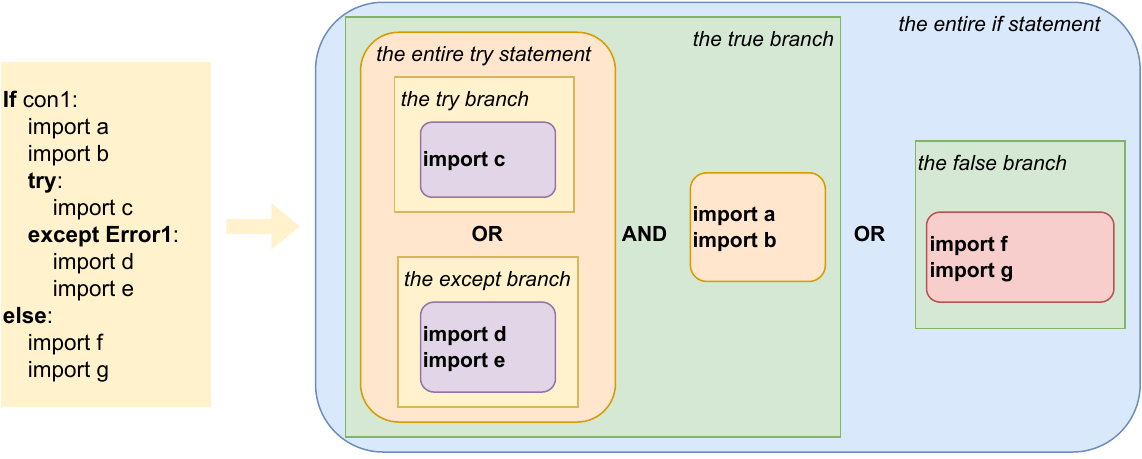}
    \caption{An example of block analysis for external imports.}
    \label{fig:blockanalysis}
\end{figure}

\textit{2) Analyze Import Blocks (STATUS 1/2 $\rightarrow$ STATUS 3).} Developers may handle different run-time environments by utilizing branch statements, such as \texttt{if-else} and \texttt{try-except}, to wrap the import statements in the code. We term this practice as \textit{multiple version control}. In such scenarios, not all imports are executed during program execution, making it essential to discern whether failures of certain imports indicate configuration issues. To address this challenge, \tool introduces import block analysis, which effectively categorizes imports under \textit{multiple version control} into import blocks. The main algorithm for import block analysis is detailed in Alg.~\ref{alg:blockanalysis}. Additionally, Fig.~\ref{fig:blockanalysis} provides an illustrative example to enhance comprehension of import block analysis.

The import block analysis takes the abstract syntax tree (AST) of the current source file as input and generates two outputs: import blocks $B$, which are sets of imports grouped based on the branch statements, and block-free imports $D$, which are import statements unaffected by any branch statements. Specifically, \tool collects all import nodes present in the AST and verifies whether they are enclosed within branch statement nodes (line 20). Import nodes not associated with branch statements are grouped as block-free imports (line 23). For import nodes associated with branch statements, \tool identifies the outermost branch statement node to facilitate further analysis (line 21). In the code of Fig.~\ref{fig:blockanalysis}, all import statements are included in a \texttt{if-else} statement, so there is no block-free import.

To accommodate nested branch statements, such as the \texttt{try-except} statement within the true branch of the \texttt{if-else} statement in Fig.~\ref{fig:blockanalysis}, \tool adopts a recursive approach (lines 1$\sim$17) to handle them, where the branch statement is divided into different blocks based on the branches (line 2). Each block is treated as a new virtual source file, and \tool recursively gathers the current import blocks and block-free imports for the given branch (lines 3$\sim$15). These current import blocks and block-free imports are then consolidated into a larger import block representative of the entire branch. This recursive process continues until all branch statements are effectively handled. The generated import blocks may exhibit nested structures due to this recursive nature. For instance, in Fig.~\ref{fig:blockanalysis}, \tool partitions the \texttt{if-else} statement into two blocks, highlighted in green. It then recursively handles statements in the two blocks. In the true branch block, \tool collects all current block-free imports and the \texttt{try-except} statement as two sub-blocks, highlighted in orange. The \texttt{try-except} block is further processed to different sub-blocks, highlighted in yellow, based on the branch \texttt{try} and \texttt{except}.

After handling all branch statements, \tool removes duplicate block-free imports and duplicate imports in the same import block. It then regards all the block-free imports $D$ as a single block and combines it with import blocks $B$ to form the final block. Therefore, there are two relationships ``AND'' and ``OR'' between blocks. The ``AND'' relationship exists between the block of block-free imports and the import blocks, indicating that both blocks will be executed in reality. The ``OR'' relationship exists between the sub-blocks inside the import block, signifying that only one sub-block will be executed. These relationships are arranged in alternating fashion at different levels of the final block. Initially, block-free imports and import blocks are distinguished, followed by further differentiation of various sub-blocks within an import block. This hierarchy allows the connection of blocks, from the outermost level to the innermost level, using a boolean expression comprising relationships of "AND$\rightarrow$OR$\rightarrow$AND$\rightarrow$OR$\rightarrow \cdot \cdot \cdot$". Therefore, \tool facilitates the validation of imports in the source file through the observation of the entire boolean expression. Upon completion of import block analysis, all library releases enter STATUS 3.

\textit{3) Execute Imports (STATUS 3 $\rightarrow$ STATUS 4).} Given the boolean expression, \tool executes the imports one by one and calculates the final value of the expression. Each successful import is regarded as \textit{True} and each failed import is regarded as \textit{False}. The value of a block with no sub-block is \textit{True} only if all the contained imports are \textit{True}. Recognizing the possibility of one run-time error masking another, \tool executes one import at a time to capture as many run-time errors as possible. Library releases whose boolean expressions for all source files are \textit{True} enter STATUS 4, indicating that they pass the \textit{Import Validation}.

\section{Experiment Setup}\label{sec:setup}

In this section, we introduce the built benchmark \bench,
the baselines in the evaluation and the experiment environment.

\textbf{Benchmark.} We include the 5,371 libraries and their 131,720 releases that pass the three checks of \tool in our benchmark \bench. As PyPI libraries themselves are Python projects and have dependencies, verified PyPI libraries can form a good benchmark to evaluate the effectiveness of automatic dependency inference approaches. We show the statistics of \bench in the last column of Table~\ref{tab:statistics}.

\textbf{Baselines.} We select three state-of-the-art automatic dependency inference approaches as our baselines:

    \textit{Pipreqs}~\cite{pipreqs}: It generates \texttt{requirements.txt} files for Python projects based on the import statements in code.
    
    \textit{Dockerizeme}~\cite{horton19dockerizeme}: It generates \texttt{Dockerfile} files for Python projects by scanning the source code. The \texttt{Dockerfile} files contain dependencies of the Python version and third-party libraries.
    
    \textit{PyEGo}~\cite{ye22knowledge}: It generates all information required to set up the run-time environments, including the Python version, the third-party libraries and system libraries. It utilizes knowledge graphs to store the information of PyPI libraries and invokes SMT solvers to solve the most proper version for each dependency.

\textbf{Metric.} We use \textbf{Pass Rate} to evaluate the performance of automatic dependency inference approaches. Pass Rate is defined as the rate of library releases whose run-time environments inferred by the approach pass all the checks of \tool.

\textbf{Environment.}
To avoid potential attacks on the host machine, \tool utilizes
Docker~\cite{docker} to install run-time environments. We re-implement all baselines using the replicate packages provided by the authors. We conduct all experiments on a Linux machine (Ubuntu 20.04 LTS) with a 112-core Intel(R) Xeon(R) Platinum 8276 CPU @ 2.20GHz and 256GB memory.

\section{Result Analysis}\label{sec:eval}

\subsection{Research Questions}

We focus on the following research questions:
\begin{itemize}
    \item \textbf{RQ1:} What are the configuration issues detected by \tool?
    \item \textbf{RQ2:} How effective are existing automatic dependency inference approaches on \bench?
\end{itemize}

To answer RQ1, we run \tool on the 8,282 libraries and their 338,069 releases, as depicted in the first column of Table~\ref{tab:statistics}, to detect configuration issues. During the \textit{Installation Check} and \textit{Import Validation}, \tool executes the libraries' code, capturing and logging run-time errors like \texttt{ImportError} encountered during the execution for analysis. In \textit{Dependency Check}, \tool collects library releases that violate the pre-defined rules in Sec.~\ref{sec:approach}. To summarize potential configuration issues, we categorize and group the reported run-time errors based on their types. We then review the error messages to identify recurring issue patterns. Regarding RQ2, due to the time-consuming nature of building run-time environments for all baselines using the complete benchmark, we opted to sample 5,000 library releases from \bench for analysis. To prevent potential bias during sampling, we initially select one release from each library, excluding a few that do not require configurations in \bench, ensuring representation from all libraries. We then randomly sample the remaining releases to reach a total of 5,000 releases in the sample dataset. 
We run the three baselines on the sampled dataset and calculate the Pass Rates of the output configurations for each baseline. Moreover, we conduct a comprehensive analysis to identify the primary reasons behind the failure of baselines to provide accurate configurations.

\subsection{RQ1: Configuration Issues}

\textbf{Overall Results on Top Popular PyPI Libraries.}
We present the statistics of libraries that successfully pass the \textit{Dependency Check} and all three checks of \tool in the second and last columns of Table~\ref{tab:statistics}, denoted as \textit{installed libraries} and \textit{validated libraries}, respectively. \textit{Installed libraries}, which are verified by \tool along with the specified run-time environments, can be correctly set up and are available to users without encountering fatal errors. We observe that there are 7,830 (95\%) installed libraries with 303,377 (90\%) releases, indicating that the setup configurations of most PyPI libraries are correct. However, the situation becomes less favorable when examining the compatibility of imports in the source code with the specified run-time environments. 
Only 5,371 (65\%) libraries, comprising 131,720 (39\%) releases, successfully pass all three checks of \tool. This indicates that approximately 30\% of libraries and 51\% of releases on the PyPI platform can be installed but may encounter source-level compatibility problems. Although these issues may not be severe enough to entirely prevent the usage of the library, they can adversely affect specific functionalities.

\begin{table*}[t]
    \centering
    \caption{Configuration issues detected by \tool. There may be multiple issues occurring in one release.}
    \scalebox{0.93}{\begin{tabular}{clcccc}
    \toprule
        \textbf{Category} & \textbf{Issue} & \textbf{Check} & \textbf{\#Releases} & \textbf{Fatal?} & \textbf{Possible Reasons} \\
    \midrule
        \multirow{5}*{\tabincell{c}{Incomplete \\ Configuration}} & Missing configuration files & Installation Check & 251 & \ding{52} & \multirow{5}*{\tabincell{c}{Missing \\required\\information}} \\
        \cmidrule{2-5}
        & Missing required libraries for setup & Installation Check & 3,318 & \ding{52} & \\
        \cmidrule{2-5}
        & Missing Python versions & Dependency Check & 55,138 & \ding{56} & \\
        \cmidrule{2-5}
        & \tabincell{l}{Missing required \\ libraries for direct imports} & Import Validation & 142,521 & \ding{56} & \\
    \midrule
        \multirow{13}*{\tabincell{c}{Incorrect \\Configuration}} & Dependency conflicts in setup & Installation Check &  6,318 & \ding{52} & Unsolvable constraints \\
        \cmidrule{2-6}
        & Incorrect Python versions & Installation Check & 4,155 & \ding{52} & Incorrect dependencies \\
        \cmidrule{2-6}
        & Other run-time Errors in setup & Installation Check & 3,464 & \ding{52} & Missing files \\ 
        \cmidrule{2-6}
        & \tabincell{l}{Inconsistent \\ configurations  with metadata} & Dependency Check & 592 & \ding{56} & Naming error \\
        \cmidrule{2-6}
        & \tabincell{l}{Inconsistent version \\ numbers with release dates} & Dependency Check & 12,018 & \ding{56} & \tabincell{l}{Confusing version orders}\\
        \cmidrule{2-6}
        & \tabincell{l}{Missing required \\ modules for indirect imports} & Import Validation & 11,023 & \ding{56} & \multirow{5}*{Incorrect dependencies} \\
        \cmidrule{2-5}
        & \tabincell{l}{Inconsistent modules in direct \\ imports with installed depenencies} & Import Validation & 6,678 & \ding{56} &  \\
        \cmidrule{2-5}
        & Other run-time Errors in imports & Import Validation & 8,178 & \ding{56} & \\
    \midrule
        \multirow{3}*{\tabincell{c}{Incorrect \\ Code}} & Missing source code & Dependency Check & 2,588 & \ding{52} &  Creating placeholders \\
        \cmidrule{2-6}
        & Parsing error & Dependency Check & 431 & \ding{52} & Invalid syntax/encoding \\
        \cmidrule{2-6}
        & Multiple version control failure & Import Validation & 15,507 & \ding{56} & Incorrect dependencies \\
    \bottomrule
    \end{tabular}}
    
    \label{tab:issues}
\end{table*}

We categorize the configuration issues identified from the library releases that failed in the three checks of \tool into three groups: \textit{Incomplete Configuration}, \textit{Incorrect Configuration}, and \textit{Incorrect Code}. All these configuration issues are presented in Table~\ref{tab:issues}. We define a configuration issue as "fatal" if it hinders the usage of the entire library release, and a configuration issue as "not fatal" if it only impacts a portion of the library's functionality. In the rest section of RQ1, we provide a detailed exploration of each configuration issue.

\textbf{Incomplete Configuration.} The issues under this category are raised due to the lack of some important information in the configurations. Specifically, the four issues are classified based on the missing information.

\textit{1) Missing configuration files.} As mentioned in Sec.~\ref{sec:background}, most libraries use configuration files such as \texttt{requirements.txt} to record the required dependencies. However, \tool identifies 251 library releases missing necessary configuration files, which directly results in failures in the \textit{Installation Check}. One such instance is the installation failure of \textit{PyAstronomy-0.10.0}, as it requires another library, \texttt{numpy}, before the successful execution of \texttt{setup.py}. However, the absence of a proper configuration file indicating the dependencies results in the failure to install the library.

\textit{2) Missing required libraries for setup.} \tool identifies 3,318 library releases that encounter installation failures due to the absence of libraries required for setup in their configurations. This configuration issue is distinguished by the occurrence of \texttt{ModuleNotFoundError} and \texttt{ImportError} during the \textit{Installation Check}. For example, in the library release \textit{translators-4.0.4}, a \texttt{ModuleNotFoundError} is triggered due to a missing module \texttt{requests}. This happens when the installer tries to obtain the version from \texttt{\_\_init\_\_.py}, but there are some external imports that are not specified in the \texttt{setup\_requires} field of \texttt{setup.py}.

\textit{3) Missing Python versions.} \tool identifies 55,138 library releases that do not indicate the required Python versions in their configurations during
\textit{Dependency Check}. The absence of specified Python versions presents significant risks to the reliability of the libraries, as the breaking changes introduced in different Python versions can impact the functionality of the libraries. A notable example is the introduction of new keywords \texttt{async} and \texttt{await} in Python version 3.5.  Identifiers \texttt{async} and \texttt{await} valid in Python versions < 3.5 become invalid in Python versions > 3.5.

\textit{4) Missing required libraries for direct imports.}
We define \textbf{direct imports} as the import statements in the source code of the current library release, and \textbf{indirect imports} as the import statements that are called by direct imports in the source code of third-party libraries required in the configurations. \tool identifies 142,521 library releases where modules required by direct imports are not installed because of missing corresponding library dependencies in the configurations. This issue is characterized by \texttt{ModuleNotFoundError} and \texttt{ImportError} occurring in direct imports in \textit{Import Validation}. For example, in the library release \textit{claripy-7.8.8.1}, there is an import statement "import celery" in the file \texttt{backends/remotetasks.py}. However, the corresponding library \textit{celery} for the module \texttt{celery} is not included in the configuration.

\finding{1}{Developers tend to provide inadequate configurations for the usage of libraries, especially for Python versions and direct imports in source code.}

\textbf{Incorrect Configuration.} The issues under this category are raised due to incorrect information in the configurations. Specifically, eight types of issues are classified based on incorrect information.

\textit{1) Dependency conflicts in setup.} Dependency conflict in the setup occurs when the dependency constraints of third-party libraries cannot be resolved to valid versions on the PyPI platform. \tool identifies 6,318 library releases with dependency conflicts during the \textit{Installation Check}, as indicated by the error message ``\textit{Could not find a version that satisfies the requirement}''. For instance, the library release \textit{accountant-0.0.6} requires \texttt{enum>=1.1.5}, but the latest version of \textit{enum} available on the PyPI platform is 0.4.7, which does not satisfy the specified constraint.

\textit{2) Incorrect Python versions.} For library releases with Python version constraints, \tool initially selects the latest Python version in the constraint for the installation and retries other Python versions in the constraints if the initial Python version fails. However, \tool finds 4,155 library releases with Python version constraints but all the Python versions in the constraints fail in \textit{Installation Check}. This suggests that the Python version constraints written by developers for these library releases are incorrect.

\textit{3) Other run-time errors in setup.} In addition to dependency conflicts and Python version issues, \tool identifies two types of run-time errors occurring during the setup process. Specifically, there are 966 library releases associated with \texttt{AttributeError} and 2,498 library releases associated with \texttt{FileNotFoundError} in the \textit{Installation Check}. The \texttt{AttributeError} is caused by incorrect setup dependencies, while the \texttt{FileNotFoundError} is a result of some non-configuration files being absent. As an example, the library release \textit{aiodocker-0.1} requires \texttt{README.md}, but it does not exist.

\textit{4) Inconsistent configurations with metadata.} \tool checks the potential inconsistencies between the configurations and the library metadata. It identifies 592 library releases with such inconsistencies. The inconsistencies primarily result from the naming errors of files or folders. For example, the metadata folder in the library release \textit{kfp-0.1.23} is named \texttt{kfp-0.1.22.dist-info}.

\textit{5) Inconsistent version numbers with release dates.} When resolving version constraints of third-party libraries, \texttt{pip} installs the latest versions that meet the constraints. The selection of the latest version is determined by comparing the version number strings. However, we have discovered cases where the version number order does not align with the release date order. For example, the library \textit{multipart} released version 2.0 in 2019 and version 0.1.1 in 2020. Developers who used this library in 2019 expected that future versions would be greater than 2.0 and thus set the constraint \texttt{multipart<0.2}. However, \texttt{pip} still considers version 0.1.1 as valid for this constraint, leading to the selection of an unexpected version. As a result, the inconsistency between the version number order and the release date order can undermine the validity of constraints set by developers. In our analysis, \tool identifies 12,018 library releases that depend on third-party libraries with this issue.

\finding{2}{Inconsistencies between version number order and release date order are prevalent in the PyPI ecosystem, undermining the validity of developers' dependency constraints.}

\textit{6) Missing required modules for indirect imports.} 
\tool identifies 11,023 library releases where modules in indirect imports are not installed, as indicated by \texttt{ModuleNotFoundError} and \texttt{ImportError} in \textit{Import Validation}. This issue arises due to two possible reasons.

Firstly, the required third-party libraries may not properly handle their own dependencies. For instance, the library release \textit{keras-bert-0.10.0} requires \textit{keras} in the configuration and has an import statement ``\texttt{import keras.backend}''. However, when importing \texttt{keras.backend}, \textit{tensorflow} is also required, but \textit{keras} does not list it as a dependency in its configuration, resulting in import failure.

Secondly, incorrect dependencies for third-party libraries in the configurations may be the cause. For example, the library release \textit{replit-1.4.0} has an external import statement ``\texttt{import flask}'', which, in turn, includes an import statement ``\texttt{from markupsafe import soft\_unicode}''. However, \texttt{soft\_unicode} is removed starting from version 2.1.0 of \texttt{markupsafe}, and there is no constraint preventing \texttt{pip} from getting the latest version of \texttt{markupsafe}, leading to the import failure.

\finding{3}{Ignoring indirect dependencies is one of the major ($\sim$ 18\%) incorrect configuration issues, indicating that developers often ignore indirect dependencies and only focus on the modules directly used in the source code. 
}

\textit{7) Inconsistent modules in direct imports with installed dependencies.} \tool identifies 6,678 library releases where modules in direct imports have corresponding library dependencies in the configurations but fail to be imported. This issue arises because \texttt{pip} automatically acquires the latest available version of the required libraries, which may lead to the exclusion of certain required modules in the direct imports if they have been removed in the latest version. A prime example of this is the library \textit{jtskit}, which is deprecated, and its developers create an empty release 0.5.0 to install another library \textit{jsontableschema}, resulting in the failure of the direct import ``import jtskit''.

\textit{8) Other run-time errors in imports.} We include all other run-time errors in this case. \tool identifies 8,178 library releases with run-time errors other than \texttt{ModuleNotFoundError} and \texttt{ImportError} in the \textit{Import Validation}, which include \texttt{TypeError}, \texttt{ValueError}, and so on.
These run-time errors are induced by the execution of global statements in the module, resulting in the failure of the module import. For instance, in the library release \textit{pandas-market-calendars-1.6.0}, there is an import statement ``import trading\_calendars''. Upon executing this import, a global statement "NP\_NAT = np.array([pd.NaT], dtype=np.int64)[0]" in the module \texttt{trading\_calendars} leads to a \texttt{TypeError} due to the use of \texttt{int()} on \texttt{NaTType} type.

\finding{4}{Developers make mistakes in writing configurations since 19\% of configuration issues are incorrect configurations. What's more, about 50\% incorrect configuration issues can only be detected by \textit{Import Validation}, indicating the importance of source-level validation.}

\textbf{Incorrect Code.} The issues under this category are raised due to the incorrect source code. Specifically, there are three cases classified based on source code errors.

\textit{1) Missing source code.} \tool identifies 2,588 library releases whose source code cannot be located in \textit{Dependency Check}. \tool cannot further validate the import statements without the source code. One possible reason we observed through manual analysis is that some library releases are published as placeholders on the PyPI platform without any actual source code. For example, the library release \texttt{mypy-protobuf-1.0} contains no source code and the \texttt{top\_level.txt} file in it indicates there is no available module in the library. 

\textit{2) Parsing error.} \tool identifies 431 library releases with Python files that cannot be parsed due to syntax errors, such as incorrect use of semicolons in Python code and encoding errors. Libraries with parsing errors cannot be handled by the Python interpreter, rendering them infeasible to be imported and used by users.

\textit{3) Multiple version control failure.} In Sec.~\ref{sec:approach}, \tool conducts import block analysis to partition import statements in different branches into separate blocks and generate boolean expressions to validate the correctness of imports. We identify 15,507 library releases whose generated boolean expressions evaluate to \textit{False} in \textit{Import Validation}, indicating that none of the branches in the branch statements successfully handle the specified run-time environments. We refrain from analyzing the run-time errors in individual branches as they may not be executed in practice. Instead, we collectively refer to these cases as "multiple version control failures," highlighting the incompatibilities between the version control in the source code and the actual run-time environments.

\finding{5}{Incorrect configurations can hardly be handled by the multiple version control logic in source code, as there are 5\% of library releases suffering from multiple version control failures.}

\subsection{RQ2: Effectiveness of Automatic Dependency Inference Approaches}

\begin{table}[t]
    \centering
    \caption{The Pass Rates (\%) of three baselines on the sampled 5,000 releases from our benchmark.}
    \begin{tabular}{cccc}
    \toprule
      \textbf{Python Version?}   &  \textbf{Pipreqs} & \textbf{Dockerizeme} & \textbf{PyEGo}\\
    \midrule
       \ding{52}  & 52.9 & 26.6 & 60.7 \\
       \ding{56} & - & 23.2 & 65.0 \\
    \bottomrule
    \end{tabular}
    
    \label{tab:baselines}
\end{table}

We evaluate the configurations provided by three baselines in two different settings, considering that the baseline Pipreqs does not output Python versions. In the first setting, we utilize the validated Python versions obtained in RQ1 and rely solely on the third-party library dependencies provided by the baselines to build run-time environments. In the second setting, we do not provide the validated Python versions and use those supplied by the baselines for building run-time environments. Table~\ref{tab:baselines} presents the Pass Rates of the three baselines under these two settings.   Notably, PyEGo achieves the highest Pass Rate of 65.0\% when using its own inferred Python versions. This suggests that approximately 35\% of library releases cannot be successfully inferred by PyEGo. On the other hand, for Pipreqs and Dockerizeme, their performance is limited, covering only 20\% to 50\% of library releases, despite the slight improvement when provided with the correct Python versions.

\begin{table}[t]
    \centering
    \caption{The issues that three baselines fail to pass the checks of \tool when we provide the Python versions. Only issues with more than 50 occurrences are included.}
    \scalebox{0.95}{
    \begin{tabular}{l|ccc}
        \toprule
       \textbf{Issue}  & \textbf{Pipreqs} & \textbf{Dockerizeme} & \textbf{PyEGo}  \\
       \midrule
        \tabincell{l}{Missing required \\ libraries for setup} & 71 & 168 & 13 \\
        \midrule
        \tabincell{l}{Missing required \\ libraries for direct imports} & 589 & 3,099 & 597 \\
        \midrule
        \tabincell{l}{Dependency \\ conflicts in setup} & 1,675 & 310 & 823 \\
        \midrule
        \tabincell{l}{Missing required modules \\ for indirect imports} & 14 & 20 & 147 \\
        \midrule
        \tabincell{l}{Multiple version \\ control failure} & 124 & 15 & 24 \\
        \bottomrule
    \end{tabular}}
    \label{tab:baselineissue}
\end{table}

To investigate the primary reasons behind the failures of the three baselines in inferring correct dependencies, we present the major issues with at least 50 occurrences (>1\%) during the check process of \tool in Table~\ref{tab:baselineissue}. Surprisingly, we find that for Pipreqs and PyEGo, approximately 68\% and 51\% of the failures, respectively, come from dependency conflicts during setup. This suggests that some dependencies provided by these baselines are not valid on the PyPI platform. Since these baselines rely on import statements to determine which libraries should be included in the configurations, there are instances where local modules share names with third-party modules or different libraries share module names, confusing their inference. In the case of Dockerizeme, around 86\% of the failures arise from missing required libraries for direct imports. This issue is also the second most common cause of failures for Pipreqs and PyEGo. One possible explanation is that the baseline databases cannot cover all libraries. For instance, PyEGo's database only includes the top 10,000 popular PyPI libraries~\cite{ye22knowledge}, while PyPI hosts over 471 thousand libraries. Regarding the other three issues in Table~\ref{tab:baselineissue}, we observe that they only frequently occur in one specific baseline, suggesting that they might arise from inappropriate designs in that particular baseline's approach.

\finding{6}{Current automatic dependency inference approaches fail to infer about 35\% of Python projects. Most failures come from dependency conflicts and the absence of required libraries in the generated configurations.}

\section{Implications}

\textbf{Fewer dependency constraints lead to more configuration issues.}
``Less is More'' seems to be a widely-used strategy to cut costs in software development. However, our findings from RQ1 reveal that 74\% of configuration issues arise from insufficient dependency constraints. While these constraints may be valid and correct during the initial release of third-party libraries, they can become outdated over time as dependencies evolve. Therefore, run-time errors may occur when using certain functionalities, which cannot be detected during the setup process of run-time environments. As a result, these issues are challenging to detect without a comprehensive evaluation of the source code. Fortunately, the resolution for these issues is relatively straightforward – by adding more strict dependency constraints. We advise third-party library developers to avoid setting open constraints like \texttt{version>1.0}. Instead, they should opt for complete and strict dependency constraints that restrict Python versions and library dependencies to the verified versions at the time of release. By doing so, developers can enhance the reliability of their libraries and mitigate potential run-time errors caused by evolving dependencies.

\textbf{Fewer conflict checks result in more dependency inference failures.}
During our analysis of why the three baselines fail to infer correct configurations, we have identified two major issues. First, some required libraries are missing, which can be resolved by updating the databases to align with the PyPI ecosystem. Second, we have observed dependency conflicts in the generated configurations. It indicates that the baselines lack sufficient conflict checks to validate the generated configurations thoroughly. For example, they do not handle the potential conflicts between the local modules inside the project and the external modules from the PyPI platform. Therefore, we recommend that future research on automatic dependency inference should incorporate more extensive conflict checks between local projects and libraries on the PyPI platform.
\section{Threats to Validity}\label{sec:threats}
The experiments and conclusions in our paper may face the following threats.

\textbf{Threats to Internal Validity.} 
During the \textit{Installation check} of \tool, we encountered some library releases that could not be installed due to timeout errors or downloading errors. These issues are primarily caused by unstable network connections. Additionally, a few cases involved extremely large libraries (>1GB) that exceeded the 600-second time limit for handling by dockers. To mitigate the impacts of this threat, we retried the installation of the library releases that failed due to network issues. For libraries encountering timeout errors, we extended the timeout limit of dockers from 600 seconds to 1,200 seconds. These methods reduce the number of failed library releases to about 22,485. However, due to limited time and competing resources, we could not address the problems for all library releases in a short timeframe. Therefore, we did not classify these library releases as having configuration issues to ensure that all configuration issues discussed in RQ1 are supported by solid and direct empirical evidence.

\textbf{Threats to External Validity.} 
During the preparation of the dataset for \tool, we select 10,000 third-party libraries from the PyPI platform. This selection is necessary as it is infeasible to check all libraries on PyPI. Additionally, we sample 5,000 library releases to evaluate the effectiveness of existing automatic dependency inference approaches. The process of data selection could potentially impact the generality of our experimental results and findings. To minimize this impact, we select the 10,000 most popular libraries that are well-maintained and widely recognized in the Python community as the dataset by following work~\cite{horton19dockerizeme,ye22knowledge,cheng22conflict}, ensuring that our study is significant as the top 10,000 libraries have a substantial influence on the PyPI ecosystem. When sampling library releases for the evaluation of current dependency inference approaches, we make the sampled releases cover all verified libraries in \bench. By doing so, we guarantee that these approaches are evaluated in diverse libraries with various functionalities rather than on different releases of the same libraries.
\section{Related Work}\label{sec:literature}

\subsection{Software Ecosystem}
For the Python software ecosystem, Valiev~\etal~\cite{valiev18ecosystem} study the ecosystem-level factors impacting the sustainability of Python projects. Bommarito~\etal~\cite{bommarito19an} conduct an empirical analysis on the Pypi ecosystem. Chen~\etal~\cite{chen20an} and Peng~\etal~\cite{peng21an} analyze the language features of Python projects. Vu~\etal~\cite{vu20typosquatting} identify the typosquatting and combosquatting attacks on the Python ecosystem. In this paper, we focus on the configuration issues in the PyPI ecosystem. For software ecosystems of other programming languages, Serebrenik~\etal~\cite{Serebrenik15challenges} study different tasks in the software ecosystem and identify six types of challenges. Mens~\cite{mens16an} study software ecosystem on the aspect of software maintenance and evolution. Lertwittayatrai~\etal~\cite{Lertwittayatrai17extracting} study the topology of the JavaScript package ecosystem. Zimmermann~\etal~\cite{Zimmermann19small} study the security threats in the npm~\cite{npm} ecosystem.  There was also a lot of work~\cite{thum14a,nadi14mining,meinicke20exploring,dubslaff22causality} studying configuration problems of software ecosystems. 

\subsection{Dependency Inference} There are a lot of efforts~\cite{huang20interactive, wang21hero} being devoted to automatically inferring environment dependencies for software. Most recently, DockerizeMe~\cite{horton19dockerizeme} infers third-party and system libraries via static analysis and dynamic analysis. V2~\cite{hortin19v2} enhances DockerizeMe and explores possible environment dependencies based on feedback-directed search. Pipreqs~\cite{pipreqs} builds the \textit{requirements.txt} files for Python projects by analyzing the \textit{import} statements in code. SnifferDog~\cite{wang21restoring} builds the execution environments for Python Jupyter notebooks. PyEGo~\cite{ye22knowledge} and PyCRE~\cite{cheng22conflict} utilize knowledge graphs to represent and analyze the dependencies between the third-party packages used by Python programs.

\subsection{Dependency Conflict Detection}
To improve the reliability of software, some researchers work on detecting potential dependency conflicts of software. Artho~\etal~\cite{artho12why} conduct a case study for conflict defects on software packages. Patra~\etal~\cite{jibesh18conflictjs} propose to detect the dependency conflicts between JavaScript libraries. Soto-Valero~\etal~\cite{sotovalero19the} study the problem of multiple versions of the same library co-existing in Maven Central. LibHarmo~\cite{huang20interactive} detects library version inconsistencies for Java Maven projects. Wang~\etal~\cite{wang18do,wang19could, wang20watchman,wang21hero, wang22will} conduct a series of empirical analyses and develop several tools to facilitate dependency conflict issue diagnosis for the ecosystem of different programming languages. These approaches focus on version-level checks while \tool conducts source-level checks by validating the import statements in source code.

There are also some research efforts on repairing dependency conflict issues. Su~\etal~\cite{su07autobash} propose to repair the inconsistencies between file systems and configuration scripts. Weiss~\etal~\cite{weiss17tortoise} capture and replay developer changes to repair the system configuration. HireBuild~\cite{hassan18hirebuild} repairs failing gradle build scripts based on the patterns from TravisTorrent dataset. SmartPip~\cite{wang22smartpip} proposes to address the efficiency problem of previous approaches on the PyPI~\cite{pypi} ecosystem.

\section{Conclusion}\label{sec:conclusion}
In this paper, we conduct an empirical study on configuration issues in the PyPI ecosystem. We propose \tool to automatically identify configuration issues in the setup stage, the packing stage and the usage stage of third-party libraries. We also build a benchmark \bench for the evaluation of automatic dependency inference approaches. We discover six findings and conclude two implications to facilitate the development of third-party libraries and future research on automatic dependency inference.

\section{Data Availability}
The proposed tool \tool and benchmark \bench are released at https://github.com/JohnnyPeng18/PyConf.

\section{Acknowledgement}

The authors would like to thank the efforts made by anonymous reviewers. The work described in this paper was supported by the Research Grants Council of the Hong Kong Special Administrative Region, China (No. CUHK 14206921 of the General Research Fund). The work was also supported by National Natural Science Foundation of China under project (No. 62002084), Natural Science Foundation of Guangdong Province (Project No. 2023A1515011959), Shenzhen Basic Research (General Project No. JCYJ20220531095214031), Shenzhen International Cooperation Project (No. GJHZ20220913143 008015), and Key Program of Fundamental Research from Shenzhen Science and Technology Innovation Commission (Project No. JCYJ20200109113403826). Any opinions, findings, and conclusions or recommendations expressed in this publication are those of the authors, and do not necessarily reflect the views of the above sponsoring entities.

\bibliographystyle{ACM-Reference-Format}
\bibliography{ref}

\end{document}